\documentstyle[12pt,psfig]{article}

\def\r{\hat\rho}
\def\rtot{\hat\rho_{\rm total}}
\def\Pke{\hat P_k^{\cal E}}
\def\Pre{\hat P_r^{\cal E}}
\def\ket#1{|#1\rangle}
\def\bra#1{\langle#1|}
\def\tr{{\rm tr}}
\def\tre{{\rm tr}_{\cal E}}
\def\Er{\hat E_r}
\def\Imin{\bar I_{\rm min}}
\def\Dsbar{\Delta\bar H}
\def\Ds{\Delta H}

\newcommand{\one}{\mbox{\tt 1}\hspace{-0.057 in}\mbox{\tt l}}
\newcommand{\binom}[2]{\left(\begin{array}{c}
                              #1\\
                              #2
                             \end{array}
                       \right)}

\begin{document}

\title{Preparation information and optimal decompositions for mixed
quantum states}

\author{
Andrei N. Soklakov\thanks{E-mail: a.soklakov@rhbnc.ac.uk} $\,$ and
R\"udiger Schack\thanks{E-mail: r.schack@rhbnc.ac.uk} \\
Department of Mathematics, Royal Holloway, \\ 
University of London, Egham, Surrey TW20 0EX, UK 
}

\date{\today}
\maketitle

\begin{abstract}
Consider a joint quantum state of a system and its environment. A measurement
on the environment induces a decomposition of the system state. Using
algorithmic information theory, we define the {\it preparation
information\/} of a pure or mixed state in a given decomposition. We
then define an {\it optimal decomposition\/} as a decomposition for
which the average preparation information is minimal. 
The average preparation information for an optimal
decomposition characterizes the system-environment correlations. We
discuss properties and applications of the concepts introduced above
and give several examples.
\end{abstract}

\section{Introduction}

It is a distinctive feature of quantum mechanics that more information
is required to prepare an ensemble of nonorthogonal quantum states
than can be recovered from the ensemble by measurements. Whereas the
von Neumann entropy of the density operator of the ensemble is bounded
above by the logarithm of the dimension of Hilbert space, $\log D$,
the preparation information for a uniform ensemble of pure states is
of the same order as $D$ \cite{Percival1992,Schack1994b,Caves1996}.

An ensemble of quantum states is defined by a list of states together
with their probabilities, $\{\r_r,p_r\}$. An ensemble can also be
regarded as a {\it decomposition\/} of the average density operator,
$\r=\sum p_r\r_r$.  Ensembles of quantum states of a system ${\cal
S}$ arise in a natural way from the correlations of ${\cal S}$ with an
environment ${\cal E}$. Given the total state $\rtot$ of the joint
system ${\cal S}\otimes{\cal E}$, any generalized measurement
or POVM \cite{Kraus1983} on ${\cal E}$ induces an ensemble
on ${\cal S}$. In this paper, we give a precise definition of the
preparation information of a state in the ensemble induced by $\rtot$
and an environment POVM.

The concept of preparation information leads naturally to a definition
of optimal ensembles or, equivalently, optimal density-operator
decompositions. The average preparation information of a state in an
optimal ensemble is then a property of $\rtot$ alone (given the split
of the total Hilbert space into ${\cal S}$ and ${\cal E}$). The average
preparation information characterizes the system-environment correlations
by the information about the environment needed to obtain a given
amount of information about the system.
 
Optimal decompositions in the sense defined here have been used for
the investigation of optimal quantum trajectories in quantum optics
\cite{Breslin1995}. Quantum trajectories are defined as follows.
In a typical quantum-optical experiment, the
system consists of selected atoms and field modes inside an optical cavity,
whereas the environment consists of the continuum of modes outside the
cavity. The time evolution of the cavity state conditional on the
results of, e.g., homodyne measurements outside the cavity defines a
quantum trajectory \cite{Carmichael,Plenio1998}. For an alternative concept
of optimality see Ref.~\cite{Wiseman1998a}. 

The average preparation information for an optimal ensemble has been proposed
as a measure of quantum chaos \cite{Schack1994b,Schack1996b}. When a chaotic
system interacts with its environment, one loses the ability to predict its
time evolution. The preparation information quantifies the amount
of information needed about the environment to keep the ability to
predict the system state to a given accuracy. In conjunction with
Landauer's principle \cite{Landauer1961}, this places a fundamental
lower limit on the free-energy cost of predicting the time evolution
of a dynamical system \cite{Caves1993b}.

The paper is organized as follows. Section~\ref{secprep} defines the
concepts of preparation information and optimal ensembles, and derives
some basic properties. In Sec.~\ref{secex}, we illustrate the theory
through	several examples. Some mathematical details are deferred to
Sec.~\ref{secdet}.

\section{Preparation information and optimal ensembles} \label{secprep}

Let $D$ and $D_{\cal E}$ denote the Hilbert-space dimensions of the
system ${\cal S}$ and the environment ${\cal E}$, respectively.  We
will normally assume that $D_{\cal E}\gg D$. Now consider a joint
state $\rtot$ on ${\cal S}\otimes{\cal E}$.  The state of the system
alone, $\r$, is then obtained by tracing out the environment,
\begin{equation}
\r=\tre(\rtot) \;.
\end{equation}
The von Neumann entropy of the system is
\begin{equation}
H = -\tr(\r\log\r) \;,
\label{eqvonNeumann}
\end{equation}
where here and throughout this paper, $\log$ denotes the base-2 logarithm.
We now perform an arbitrary measurement on the environment \cite{Kraus1983}, 
described by a POVM, $\{\Er\}$, where the $\Er$ are positive
environment operators such that
\begin{equation}
\sum_r\Er=\one_{\cal E}=\mbox{(environment unit operator).}
\end{equation}
The probability of obtaining result $r$ is given by
\begin{equation}
p_r= \tr(\rtot\Er)\;,
\end{equation}
and the system state after a measurement that yields result $r$ is
\begin{equation}
\r_r=
{\tre(\rtot\Er)\over p_r}\;.
\end{equation}
By summing over $r$ and using the completeness of the POVM, we obtain
\begin{equation}
\sum_rp_r\r_r=\tre(\rtot)
=\r\;.
\end{equation}
The ensemble $\{\r_r,p_r\}$ forms a decomposition of $\r$.  To
characterize the ensemble, we define the system entropy conditional on
measurement outcome $r$,
\begin{equation}
H_r = -\tr(\r_r\log\r_r) \;,
\end{equation}
the average conditional entropy,
\begin{equation}
\bar H = \sum_r p_r H_r \;,
\end{equation}
and the average entropy decrease due to the measurement, $\Dsbar=H-\bar H$.
These quantities obey the double inequality
\begin{equation}
0 \le \Dsbar \le H \;,
\end{equation}
which follows from the concavity of the von Neumann entropy
\cite{Balian1991,Caves1994a}. The content of the first of these
inequalities is that a measurement on the environment will not, on the
average, increase the system entropy.

Now let $\{\r_r,p_r\}$ be the ensemble induced by the POVM
$\{\Er\}$. We denote by $I(\r_k|\rtot,\{\Er\})$ the conditional algorithmic
information to specify $\r_k$, given the ensemble (see
\cite{Chaitin1987a} and references in \cite{Schack1997a}). The
quantity $I(\r_k|\rtot,\{\Er\})$ defines the {\it preparation
information\/} of the state $\r_k$, given the total state $\rtot$ and
the POVM. We also define the {\it average preparation
information\/}
\begin{equation}
\bar I(\rtot,\{\Er\}) = -\sum_r p_r \log p_r \;.
\end{equation}
This definition is justified, because the average algorithmic information
can be bounded above and below as follows: \cite{Schack1997a}
\begin{equation}
-\sum_r p_r \log p_r \le \sum_k p_k I(\r_k|\rtot,\{\Er\}) \le
-\sum_r p_r \log p_r + 1 \;.
\end{equation}
The average preparation information is never smaller than the average system
entropy decrease $\Dsbar$, 
\begin{equation}
\bar I(\rtot,\{\Er\}) \ge \Dsbar \;.
\end{equation}
This inequality is a consequence of a general theorem about average
density operators \cite{Balian1991,Caves1994a}.

The next step is to define a {\it $\Ds$-decomposition\/} of $\r$ as a
decomposition for which $\Dsbar\ge\Ds$, and an {\it optimal
$\Ds$-decomposition\/} of $\r$ as a $\Ds$-decomposition with minimal
average preparation information $\bar I$. The average
preparation information for an optimal $\Ds$-decomposition,
\begin{equation}
\Imin = \inf_{\Ds-{\rm decompositions}} \bar I \;,
\label{eqdefimin}
\end{equation}
is then a property of $\rtot$, and characterizes the
system-environment correlations.  (If there is no $\Ds$ decomposition
for which $\bar I$ is minimal, we will call any decomposition optimal
for which $\bar I<\Imin+\epsilon$ for some given small constant
$\epsilon$.)   The quantity $\Imin$ is the
information about the environment needed to reduce the system entropy
by $\Ds$. A useful generalization results from taking the infimum in
Eq.~(\ref{eqdefimin}) over a restricted class of POVMs, as in
the quantum-optical example of Ref.~\cite{Breslin1995}. This defines
ensembles that are optimal with respect to a given class of
environment measurements.

\section{Examples}       \label{secex}

In this section, $\{\Pke, k=1,\ldots,D_{\cal E}\}$ denotes a complete
set of orthogonal environment projectors. In the three examples
discussed below, we will restrict the class of environment
measurements to orthogonal projections of the form 
\begin{equation}
\Er=\sum_{k\in K_r}\Pke \;,
\label{eqgroupingPOVM}
\end{equation}
where $K_r\subset\{1,\ldots,D_{\cal E}\}$. In all three
examples, it seems intuitively clear that ensembles which are optimal
with respect to this class of measurements are also, to a good
approximation, optimal with respect to the class of all possible
environment measurements.  We have not, however, been able to prove this
statement rigorously.

\subsection{A trivial example}

Here, the system is a qubit, for which the dimension of Hilbert space
is $D=2$.  Let $\ket0$ and $\ket1$ be orthogonal basis states for the
qubit, define $\ket{\psi_1}=\ket0$, $\ket{\psi_2}=\ket1$,
$\ket{\psi_3}={1\over\sqrt2}(\ket0+\ket1)$ and
$\ket{\psi_4}={1\over\sqrt2}(\ket0-\ket1)$, and let
\begin{equation}
\rtot={1\over4}\sum_{k=1}^4 \ket{\psi_k}\bra{\psi_k} \otimes \Pke 
\end{equation}
be the joint density operator of system and environment. The state of
the system alone is then given by 
\begin{equation}
\r = \tr_{\cal E}(\rtot) = {1\over2} (\ket0\bra0+\ket1\bra1) \;, 
\end{equation}
for which the system entropy in the absence of measurements is given
by $H=1$.

Suppose we want to reduce the system entropy by $\Ds=1$bit, i.e., we
want the conditional system state to be pure. The only environment
measurement achieving this is given by $\hat E_r=\Pre$, which results
in the unique and therefore optimal $\Ds=1$ ensemble given by
$\r_r=\ket{\psi_r}\bra{\psi_r}$, $p_r=1/4$. The average preparation
information for this ensemble is $\bar I=2$bits, and hence $\Imin=2$bits.

For a different value of $\Ds$, consider the ensemble defined by
$\r_1= {1\over2} (\ket{\psi_1}\bra{\psi_1}+\ket{\psi_3}\bra{\psi_3})$,
$\r_2= {1\over2} (\ket{\psi_2}\bra{\psi_2}+\ket{\psi_4}\bra{\psi_4})$,
$p_1=p_2=1/2$, which is induced by the POVM $\hat E_1=\hat P_1^{\cal
E}+\hat P_3^{\cal E}$ and $\hat E_2=\hat P_2^{\cal E}+\hat P_4^{\cal
E}$. For this ensemble,
$H_1=H_2=-\tr(\r_1\log\r_1)\simeq0.81$, and hence 
$\Dsbar=H-H_1\simeq0.19$. The average preparation information is 
$\bar I=1$. It is
easy to see that, with respect to our restricted class of
measurements, this is an optimal $\Dsbar$-decomposition, and hence
$\Imin=1$. In this example, to obtain 0.19 bits of information about
the system, 1 bit of information about the environment is needed.

\subsection{Random vectors in Hilbert space} \label{subsecrand}

In the trivial example considered above, the average
preparation information $\Imin$ for an optimal ensemble is significantly larger
than the corresponding entropy reduction $\Dsbar$. In the present
subsection, we show that $\Imin$ can vastly exceed $\Dsbar$.

Assume that $\log D_{\cal E}\gg\log D$ and consider
\begin{equation}
\rtot={1\over D_{\cal E}}\sum_{k=1}^{D_{\cal E}} 
\ket{\psi_k}\bra{\psi_k} \otimes \Pke \;,
\end{equation}
where the $\ket{\psi_k}$ are distributed randomly in $D$-dimensional
(projective) Hilbert space \cite{Wootters1990}. Here, the system
entropy in the absence of measurements is $H\simeq\log D$.  It has been
conjectured \cite{Schack1996b,Schack1996a} that states of a similar
form arise from the interaction of a chaotic system with a random
environment. We will see that the complexity of the resulting
system-environment correlations, as quantified by the average
preparation information, is very large. This is in marked contrast to
the third example discussed below.

Environment measurements of the form (\ref{eqgroupingPOVM}) correspond
to grouping the vectors $\ket{\psi_k}$ into disjoint groups. We construct an
approximation to an optimal measurement by grouping the vectors into
Hilbert-space spheres of radius $\phi$. (See Ref.~\cite{Schack1996b}
for a detailed argument.)  We assume that $D_{\cal E}$ is
sufficiently large so that the state vectors in each such sphere fill
it randomly. Since all spheres are chosen to be of equal size, the
average entropy $\bar H$ is equal to the entropy of one sphere, i.e.,
the entropy of a uniform mixture of states within a Hilbert-space
sphere of radius $\phi$, given by \cite{Schack1994b}
\begin{equation}
\bar H = -\left(1-\frac{D-1}{D}\sin^2\phi\right)
\log\left(1-\frac{D-1}{D}\sin^2\phi\right)
\;-\frac{D-1}{D}\sin^2\phi\,
\log\left(\frac{\sin^2\phi}{D}\right) \;.
\label{eqhphi}
\end{equation}
The volume contained within a sphere of radius $\phi$ in Hilbert space
is $(\sin\phi)^{2(D-1)}V_D$, where $V_D$ is the total volume of
projective Hilbert space \cite{Schack1994b}. The number of spheres of
radius $\phi$ in $D$-dimensional Hilbert space is thus
$(\sin\phi)^{-2(D-1)}$, so the information needed to specify a
particular sphere is
\begin{equation}
\Imin \simeq \tilde I_{\rm min} \stackrel{\rm def}{=}
  \log\left( (\sin\phi)^{-2(D-1)} \right)
 = -(D-1) \log(\sin^2\phi) \;.
\label{eqiphi}
\end{equation}
The information $\tilde I_{\rm min}$ slightly underestimates the
actual value of $\Imin$, because the perfect grouping
into nonoverlapping spheres of the same size assumed by
Eq.~(\ref{eqiphi}) does not exist.

As an example, let us choose a Hilbert-space dimension $D=101$ and a radius of
$\phi\simeq1.025$. Equations (\ref{eqhphi},\ref{eqiphi}) then give
$\Dsbar=\log D-\bar H\simeq1$ and $\tilde I_{\rm min}\simeq45.3$, which means
that here, to
obtain 1 bit of information about the system, more than 45 bits of information
about the environment are needed.

Using Eq.~(\ref{eqiphi}) to eliminate $\phi$ from Eq.~(\ref{eqhphi})
gives a complicated expression for $\tilde I_{\rm min}$ as a function
of $\Dsbar$ \cite{Schack1997c}, which is a good approximation to the
average preparation information for an optimal $\Dsbar$ ensemble. Figure 1
shows this function for a Hilbert space dimension $D=101$. To obtain
more insight into the properties of this curve, we consider the
derivative \cite{Schack1997c}
\begin{equation}
{d\tilde I_{\rm min} \over d\Dsbar} = 
 {D \over \sin^2\phi \ln(1+D\cot^2\phi)} \;,
\label{eqdi}
\end{equation}
which is the marginal tradeoff between between information and entropy. For
$\phi$ near $\pi/2$, so that $\epsilon=\pi/2-\phi\ll1$, the information becomes
$\tilde I_{\rm min}=(D-1)\epsilon^2/\ln2$, and the derivative
(\ref{eqdi}) can be written as 
\begin{equation}
{d\tilde I_{\rm min} \over d\Dsbar} \simeq {D \over \ln(1+D\epsilon^2)} \;,
\label{eqdiapprox}
\end{equation}
which is proportional to $D$ with a slowly varying logarithmic correction.
We have thus identified a situation where the average preparation
information is of the same order as the dimension of Hilbert space $D$,
despite the fact that the von Neumann entropy of a state cannot exceed
$\log D$.

\subsection{Random coherent states} \label{subseccoherent}

In this example the system considered is a spin-$j$ particle, for
which the dimension of Hilbert space is $D=2j+1$.  As in the preceding
section, we assume that the Hilbert-space
dimension of the environment is much larger than $D$, $D_{\cal E}\gg D$, 
and consider
\begin{equation}
\rtot={1\over D_{\cal E}}\sum_{k=1}^{D_{\cal E}} 
\ket{\psi_k}\bra{\psi_k} \otimes \Pke \;,
\end{equation}
but now we choose the $\ket{\psi_k}$ to be distributed randomly on the
submanifold of angular-momentum coherent states (\ref{defrhocoh}). 
We will see that
the resulting complexity of the system-environment correlations, 
as quantified by the average preparation information, is small.

The angular momentum coherent state $|\theta,\phi\rangle$ can be
defined by rotating the $\hat J_z$ eigenstate $|j;j\rangle$ through
Euler angles $\phi$ around the $z$-axis, and then by  $\theta$ around the new
$y$-axis. This gives \cite{Peres}
\begin{equation}                                            \label{defcoherent}
|\theta,\phi\rangle=\sum_{m=-j}^{j}|j;m\rangle\binom{2j}{j+m}^{\frac{1}{2}}
\cos^{j+m}(\theta/2)\sin^{j-m}(\theta/2)e^{-im\phi}\;.
\end{equation}
Each coherent states corresponds to a point on the surface of a
three-dimensional sphere. Assuming that $D_{\cal E}$ is sufficiently
large the state of the system alone is
\begin{equation}                       \label{defrhocoh}
\hat{\rho}=\tr_{\cal E}(\rtot)=
\frac{1}{4\pi}\int_{0}^{2\pi}\int_{0}^{\pi} 
 |\theta,\phi\rangle\langle\theta,\phi|\sin\theta\,d\theta\,d\phi\;.
\end{equation}

As in the previous section, environment measurements of the form
(\ref{eqgroupingPOVM}) correspond to grouping the vectors
$\ket{\psi_k}$ into disjoint groups. Approximately optimal
measurements correspond to grouping the vectors into approximately
equal, compact areas on the surface of the sphere. We choose the areas to be of
the form 
\begin{equation}
\Omega_r(\Theta)=\{\theta,\phi: \arccos[\underbar{n}(\theta,\phi)
\underbar{n}(\theta_r,\phi_r)]\leq\Theta\}
\end{equation}
centered at points $(\theta_r,\phi_r)$.
The corresponding density operators 
\begin{equation}
\hat{\rho}_r(\Theta)=\frac{\int_{\Omega_r(\Theta)}
            |\theta,\phi\rangle\langle\theta,\phi|\sin\theta\,d\theta\,d\phi}{
             \int_{\Omega_r(\Theta)}\sin\theta\,d\theta\,d\phi}\;,
\label{defrhoTheta}
\end{equation}
can be used to construct a nearly optimal decomposition of $\hat{\rho}$.
The preparation information $\Imin$ is then
approximately given by
\begin{equation}
\Imin\simeq\tilde{I}_{\rm min}\stackrel{\rm def}{=}
\log \frac{4\pi}{2\pi(1-\cos\Theta)}\;, 
\label{eqImincoh}
\end{equation}
where the denominator is the area of $\Omega_r(\Theta)$
 
In the following section, we show that $\hat{\rho}_r(\Theta)$,
in the coordinates where $(\theta_r,\phi_r)=(0,0)$, 
can be written
in the diagonal form
\begin{equation}
\hat{\rho}_r(\Theta)=\sum_{m=-j}^{j}|j;m\rangle\lambda^{\Theta}_m\langle j;m|\;,
\label{general}
\end{equation}
where
\begin{equation}                                              
\lambda^\Theta_m=
\frac{(2j)!\;\sin^{2(j-m)}\frac{\Theta}{2}}{(j+m)!\;(j-m+1)!}
             F(-j-m,j-m+1,j-m+2,\sin^2\frac{\Theta}{2})\;,
\label{generaleigen}
\end{equation}
and where $F$ is the hypergeometric function ${}_2F_1$.
Since all density operators $\r_r(\Theta)$ in the decomposition
of $\hat{\rho}$ have the same entropy, the average entropy
$\bar H$ can be written as
\begin{equation}
\bar H =-\sum_{m=-j}^{j} \lambda^\Theta_m\log\lambda^\Theta_m\;.
\label{eqentrocoh}
\end{equation}
For the entropy of the system, Eq.~(\ref{eqvonNeumann}), in the absence of
measurements, $\hat{\rho}=\hat{\rho}_r(\pi)$, we have
(see Eq.~\ref{GradshteynRyzhik})
\begin{equation}
 H= -\sum_{m=-j}^{j} \lambda^\pi_m\log\lambda^\pi_m =\log(2j+1)=\log D\;.
\end{equation}
We can analyse $\bar{H}$ in detail for the important
special value $\Theta=\pi/2$, for which  the measurement corresponds to
a grouping into two disjoint hemispheres, and is therefore strictly
optimal. For such a grouping, $\Imin=1$.
In the next section, we show that the eigenvalues  
$\lambda_m^{\pi/2}$
obey the bounds
\begin{eqnarray}
0\leq\lambda_m^{\pi/2}\leq\frac{1}{j}e^{-\sqrt[3]{j}/3}  \;, &&
                            m<-1-j^{2/3} \;, \cr
\frac{2}{2j+1}(1-e^{-\sqrt[3]{j}/3}-4^{-j})\leq\lambda_m^{\pi/2}\leq\frac{1}{j}
  \;, &&
                             m>1+j^{2/3}\;.
\label{eqj23}
\end{eqnarray}
Using these bounds we have derived the following asymptotic expression
for the average entropy:
\begin{equation}
\bar H= - \sum_{m=-j}^{j}\lambda_m^{\pi/2}\log\lambda_m^{\pi/2}=
\log j + O(\sqrt[3]{j}e^{-\sqrt[3]{j}/3}) \;,
\end{equation}
and hence, in the limit $j\rightarrow\infty$, 
\begin{equation}
{\Imin\over\Dsbar} = {1\over \log(2j+1)-\bar H} \longrightarrow 1\;.
\end{equation}

Figure 2 shows a parametric plot of the average preparation
information $\tilde I_{\rm min}$ versus the average entropy reduction
$\Dsbar=H-\bar H$ for $j=50$, i.e., D=101. It can be seen that
for moderate values of $\Dsbar$, $\tilde I_{\rm min}\simeq\Dsbar$. To
reduce the system entropy by 1 bit, not more than approximately 1 bit
of information about the environment is needed. This should be
compared to the previous example, where the required environment
information is of the same order as $D$. In the limit of $\Dsbar$
approaching its maximum value $H=\log D$, the information $\tilde I_{\rm
min}$ diverges. This is due to the fact that an infinite amount of
information is needed to specify a general state exactly. The
complexity of the system-environment correlations is characterized by
the slope of the curve for small values of $\Dsbar$ rather than its
asymptotic behaviour for $\Dsbar\rightarrow\log D$.

\section{Mathematical details}  \label{secdet}

Our task is to calculate the eigenvalues of $\hat{\rho}_r(\Theta)$ given by
Eq.~(\ref{generaleigen}) and to derive the expression (\ref{eqj23}) for the
case $\Theta=\pi/2$.
Choosing the
coordinate system such that $(\theta_r,\phi_r)=(0,0)$, we have
\begin{equation}
\hat{\rho}_r(\Theta)=\frac{\int_{0}^{2\pi}\int_{0}^{\Theta}
            |\theta,\phi\rangle\langle\theta,\phi|\sin\theta\,d\theta\,d\phi}{
             2\pi(1-\cos\Theta)}\;.
\end{equation}
Substitution of (\ref{defcoherent}) and subsequent integration over $\phi$
gives
\begin{equation}                                               \label{rhoTheta}
\hat{\rho}_r(\Theta)=\frac{4}{1-\cos\Theta}
             \sum_{m=-j}^{j}|j;m\rangle\langle j;m|\binom{2j}{j+m}
             \Lambda^\Theta(j+m,j-m)\;,
\end{equation}
where
\begin{equation}                                                 \label{Lambda}
\Lambda^\Theta(p,q)=
            \int_{0}^{\Theta/2}\cos^{2p+1}\vartheta\sin^{2q+1}
            \vartheta\,d \vartheta\;.
\end{equation}
Since $\hat{\rho}_r(\Theta)$ is diagonal in the $|j;m\rangle$ basis, 
the task of finding the eigenvalues is equivalent to the problem of
evaluating the integral $\Lambda^\Theta(p,q)$. Consider the integrand
\begin{equation}
\cos^{2p+1}\vartheta\sin^{2q+1}\vartheta=
     -\frac{1}{2}\cos^{2p}\vartheta(1-\cos^2\vartheta)^q
        \frac{d\cos^2\vartheta}{d\vartheta}\;.
\end{equation}
Using the formula \cite{GradshteynRyzhik}
\begin{equation}   \label{onepluszton}
(1+z)^n=F(-n,b;b;-z) \;,
\end{equation}
where $F$ is the hypergeometric function ${}_2F_1$, we obtain
\begin{equation}                                             \label{coherent8}
\cos^{2p+1}\vartheta\sin^{2q+1}\vartheta=
 -\frac{1}{2}(\cos^2\vartheta)^pF(-q,b;b;\cos^2\vartheta)
              \frac{d\cos^2\vartheta}{d\vartheta}\;.
\end{equation}
We now use \cite{AbramowitzStegun}
\begin{equation}
\frac{d^n}{dz^n}[z^{c-1}F(a,b;c;z)]=
\frac{\Gamma(c)}{\Gamma(c-n)}z^{c-n-1}F(a,b;c-n;z),
\end{equation}
with $a=-q$, $c=b+1$ and $z=\cos^2\vartheta$ to get
\begin{equation}                                            \label{coherent10}
b(\cos^2\vartheta)^{b-1}F(-q,b;b;\cos^2\vartheta)
  =\frac{d}{d\cos^2\vartheta}[(\cos^2\vartheta)^bF(-q,b;b+1;\cos^2\vartheta)].
\end{equation}
Comparing (\ref{coherent8}) and (\ref{coherent10}) we see that, choosing
$b=p+1$, we find
\begin{equation}
\cos^{2p+1}\vartheta\sin^{2q+1}\vartheta
=-\frac{1}{2(p+1)}\frac{d}{d\vartheta}
  [\cos^{2p+2}\vartheta F(-q,p+1;p+2;\cos^2\vartheta)]
                          \;,
\end{equation}
and hence, using \cite{GradshteynRyzhik}
\begin{equation} \label{GradshteynRyzhik}
F(a,b;c;1)=\frac{\Gamma(c)\Gamma(c-a-b)}{\Gamma(c-a)\Gamma(c-b)}
  \ \;, \ \ \ \ \Re c > \Re (a+b) \;,
\end{equation}
one obtains
\begin{equation}                                   \label{AlmostThere}
\Lambda^\Theta(p,q)=\frac{p!\,q!}{2\cdot(p+q+1)!}
    -\frac{\cos^{2p+2}\frac{\Theta}{2}}{2(p+1)}
    F(-q,p+1;p+2;\cos^2\frac{\Theta}{2})\;,\ \ \ q >-1\;.
\end{equation}
To simplify the above equation we return to the definition
(\ref{Lambda}) and split the integration to get
\begin{equation}
\Lambda^{\Theta}(p,q)
       =\int_0^{\frac{\pi}{2}}\cos^{2p+1}\vartheta \sin^{2q+1}\vartheta\,
                          d\vartheta
                              +
\int_{\frac{\pi}{2}}^{\frac{\Theta}{2}}
             \cos^{2p+1}\vartheta \sin^{2q+1}\vartheta\,
                          d\vartheta\;.
\end{equation}
The first integral is proportional to the beta function and the second
integral can be transformed into $\Lambda^{\pi-\Theta}(q,p)$ by substitution
$\vartheta\to\pi/2-\vartheta$ so that
\begin{equation} \label{littletrick}
\Lambda^{\Theta}(p,q)= \frac{p!\,q!}{2\cdot(p+q+1)!}
                       -\Lambda^{\pi-\Theta}(q,p)\;.
\end{equation}
Using this formula Eq.~(\ref{AlmostThere}) can be transformed into
\begin{equation}
\Lambda^{\Theta}(p,q)= \frac{\sin^{2q+2}{\Theta \over 2}}{2(q+1)}
    F(-p,q+1;q+2;\sin^2{\Theta \over 2})\;.
\end{equation}
Expression (\ref{rhoTheta}) for $\hat{\rho}_r(\Theta)$ can now be rewritten
in the compact form of Eq.~(\ref{general}).

To investigate the case $\Theta=\pi/2$, we
calculate directly
\begin{equation}                                                \label{Lambda1}
\Lambda^{\frac{\pi}{2}}(p,q)=
    \int_0^{\frac{\pi}{4}}\cos^{2p+1}\vartheta \sin^{2q+1}\vartheta\,
                          d\vartheta\;.
\end{equation}
Substitution of $t=\tan^2\vartheta$ gives
\begin{equation}
\Lambda^{\frac{\pi}{2}}(p,q)=\frac{1}{2}\int_{0}^{1}t^q(1+t)^{-(p+q+2)}\,dt.
\end{equation}
Using the integral representation of the hypergeometric function
\cite{AbramowitzStegun},
\begin{equation} \label{IntRepresentF}
F(a,b;c;z)=\frac{1}{B(b,c-b)}
\int_{0}^{1}t^{b-1}(1-t)^{c-b-1}(1-tz)^{-a}\,dt\ \;, \;\; 
\Re c > \Re b>0\;,
\end{equation}
we find
\begin{equation}                                            \label{CloseLambda}
\Lambda^{\frac{\pi}{2}}(p,q)=\frac{F(p+q+2,q+1;q+2;-1)}{2(q+1)}\;,\;q>-1\;,
\end{equation}
which is a rather compact expression. We can obtain additional
insight in the following way. Consider the
Gauss formula for the so-called contiguous functions $F(a,b-1;c;z)$
and $F(a,b;c+1;z)$ \cite{AbramowitzStegun}
\begin{equation}
c(1-z)F(a,b;c;z)-cF(a,b-1;c;z)+(c-a)zF(a,b;c+1;z)=0 \;.
\end{equation}
For the case $b=c$ and $z=-1$ we get using (\ref{onepluszton})
\begin{equation} \label{Gauss}
F(a,b;b+1;-1)=\alpha(b)F(a,b-1;b;-1)+\beta(b),
\end{equation}
where
\begin{equation}
\alpha(b)=\frac{b}{a-b}\;,{\mbox\ \ \ \ }\beta(b)=-2^{1-a}\alpha(b)\;.
\end{equation}
Iteration of (\ref{Gauss}) $b-1$ times gives
\begin{equation}
F(a,b;b+1;-1)=F(a,1;2;-1)\prod_{k=0}^{b-2}\alpha(b-k)
                                             +\sum_{s=0}^{b-2}\alpha(b-s)
                                             \prod_{k=0}^{s-1}\beta(b-k)\;.
\end{equation}
Noticing that
\begin{equation}
\prod_{k=0}^{s}\alpha(b-k)=\frac{b!\,(a-b-1)!}{(b-s-1)!\,(a-b+s)!}
\end{equation}
and substituting $x=b-1-s$, we obtain
\begin{equation}
F(a,b;b+1;-1)=\frac{b!\,(a-b-1)!}{(a-1)!}[(a-1)F(a,1;2;-1)
           - 2^{1-a}\sum_{x=1}^{b-1}\binom{a-1}{x}]\;.
\end{equation}
The value of $F(a,1;2;-1)$ can be calculated using the integral
representation (\ref{IntRepresentF})
\begin{equation}
F(a,1;2;-1)=\frac{1-2^{1-a}}{a-1}\;,
\end{equation}
and hence we find
\begin{equation}
F(a,b;b+1;-1)=\frac{b!\,(a-b-1)!}{(a-1)!}
              [1-2^{1-a}\sum_{x=0}^{b-1}\binom{a-1}{x}]\;.
\end{equation}
Using this formula, Eq.~(\ref{CloseLambda}) can be rewritten as
\begin{equation}                                         \label{AlmostThere1}
\Lambda^{\frac{\pi}{2}}(p,q)=\frac{1}{2(p+q+1)}\binom{p+q}{p}^{-1}
                [1-2^{-(p+q+1)}\sum_{x=0}^{q}\binom{p+q+1}{x}]\;.
\end{equation}
Using (\ref{littletrick}) we have
\begin{equation}
\Lambda^{\frac{\pi}{2}}(p,q)
    =\frac{2^{-(p+q+2)}}{p+q+1}\sum_{x=0}^{p}\binom{p+q+1}{x}\;,
\end{equation}
and therefore
\begin{equation}
\lambda^{\frac{\pi}{2}}_m
 =\frac{2}{2j+1}\sum_{x=0}^{j+m}\binom{2j+1}{x}2^{-(2j+1)}\;.
\end{equation}
The sum on the right hand side is the
univariate cumulative distribution function \cite{AbramowitzStegun}
\begin{equation}
G(y;n,p)=\sum_{x=0}^{y}P(x;n,p)
\end{equation}
for the binomial distribution
\begin{equation}
P(x;n,p)=\binom{n}{x}p^x(1-p)^{n-x}\;,
\end{equation}
where $n=2j+1$, $p=1/2$ and $y=j+m$. In the limit $j\to\infty$, $G(y;n,p)$
approaches the step function
\begin{equation}                                           \label{StepFunction}
 \lim_{j\to\infty}G(j+m;2j+1,1/2)= \left\{\begin{array}{ll}
                                   0\;, & \lim_{j\to\infty}\frac{m}{j}< 0\;, \\
                                       \frac{1}{2}\;, & m=0\;, \\
                                   1\;, & \lim_{j\to\infty}\frac{m}{j}> 0\;.
                                       \end{array}\right.
\end{equation}
To obtain the behaviour for large but finite values of $j$, we use the
Chernoff bounds \cite{Chernoff}
\begin{eqnarray}
\sum_{x<np(1-\epsilon)}P(x;n,p)&\leq& e^{-\epsilon^2np/3} \;,\nonumber\\
\sum_{x>np(1+\epsilon)}P(x;n,p)&\leq& e^{-\epsilon^2np/3} \;,
\end{eqnarray}
which are valid for $0\le\epsilon\le1$. Choosing
$\epsilon=j^{-1/3}$,
we obtain Eq.~(\ref{eqj23}), 
in agreement with Eq.~(\ref{StepFunction}).

\section{Acknowledgements}

The authors profited from discussions with Jens G.\ Jensen. This work
was supported in part by the UK Engineering and Physical Sciences
Research Council.

%\bibliographystyle{acm}
%\bibliography{/home/rschack/lit/p}

\newpage

\begin{figure}
\psfig{figure=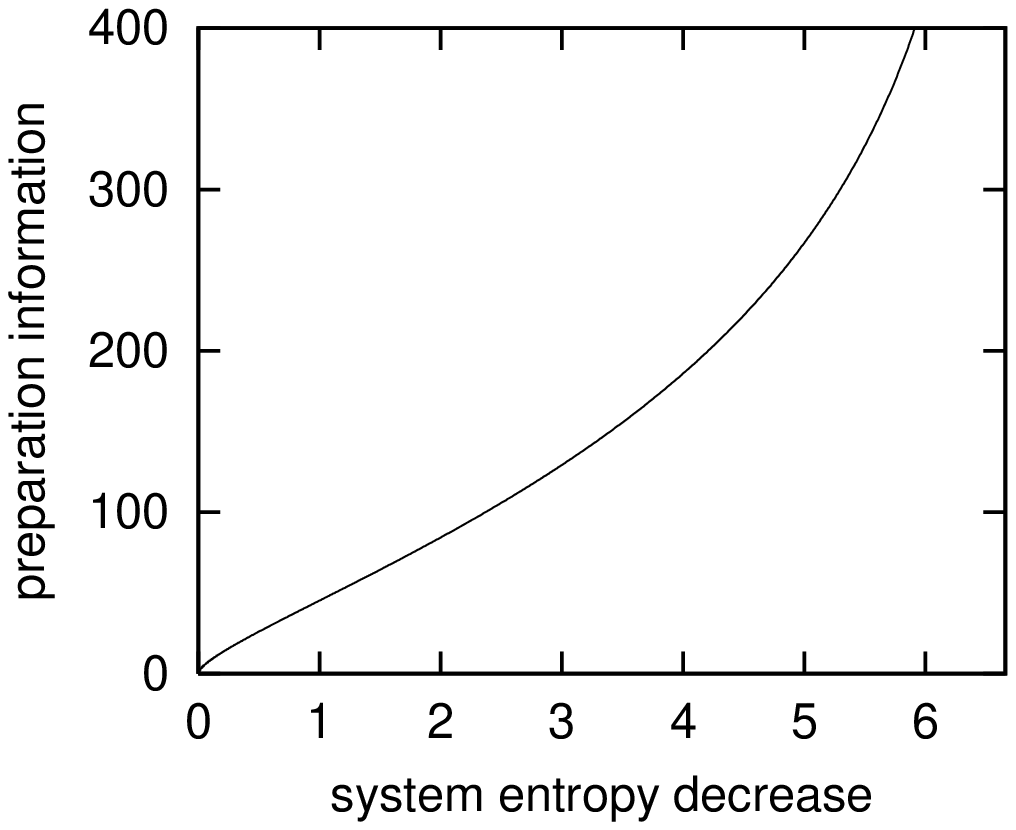}
\caption{Average preparation information $\tilde I_{\rm min}$ in bits,
versus average entropy reduction $\Dsbar$ in bits, for the example of
subsection~\protect\ref{subsecrand}.}
\label{fig1}
\end{figure}

\begin{figure}
\psfig{figure=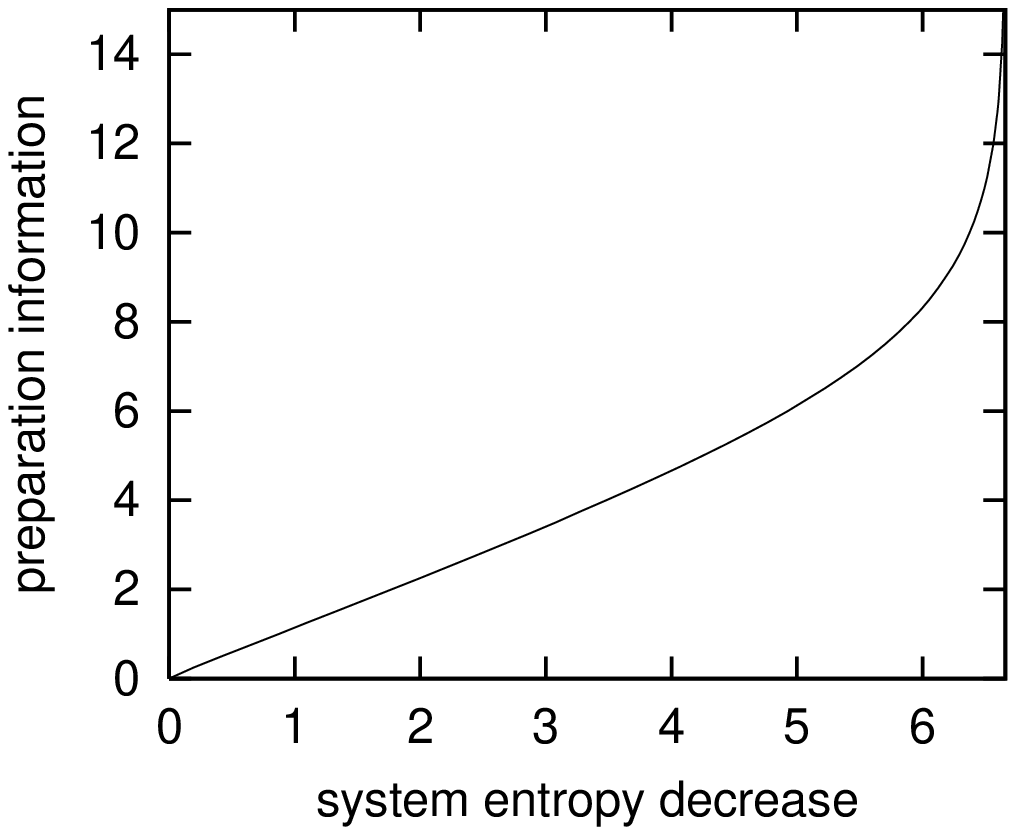}
\caption{Average preparation information $\tilde I_{\rm min}$ in bits,
versus average entropy reduction $\Dsbar$ in bits, for the example of
subsection~\protect\ref{subseccoherent}.}
\label{fig2}
\end{figure}

\end{document}